\documentclass[a4paper, 12pt]{article} 
\usepackage[right=1in,left=1in,top=1in,bottom=1in]{geometry}
\usepackage{hyperref}
\usepackage{amsmath}
\usepackage[utf8]{inputenc}
\usepackage[T1]{fontenc} 
\usepackage{pxfonts}
\usepackage{tikz}   
\usepackage{tikz-3dplot}
\usepackage{tcolorbox}
\usepackage{pgfplots}
\usepackage[titles]{tocloft}

\usepackage{amsthm}
\newcommand{\x}[1]{{#1}}
\newtheorem{theorem}{Theorem}

\title{\Large{Algorithmic Randomness, Exchangeability, and the Principal Principle}} 

\author{Jeffrey A. Barrett\footnote{Department of Logic and Philosophy of Science, University of California, Irvine, Irvine, CA 92697-5100. Email: j.barrett@uci.edu} 
~~and 
Eddy Keming Chen\thanks{Hal{\i}c{\i}o{\u g}lu Data Science Institute and Department of Philosophy, University of California, San Diego, 9500 Gilman Dr, La Jolla, CA 92093-0119. Email: eddykemingchen@ucsd.edu}}

\date{Accepted version, \textit{The British Journal for the Philosophy of Science}}

\begin{document}

\maketitle

\begin{abstract}

We introduce a framework uniting algorithmic randomness with exchangeable credences to address foundational questions in philosophy of probability and philosophy of science. As an application, we show how one might derive the Principal Principle---the norm that rational credence should match known objective chance---in the context of this framework without circularity. The derivation brings together de Finetti's exchangeability, Martin-L\"of randomness, Lewis's and Skyrms's chance-credence norms, and statistical constraining laws (\href{https://arxiv.org/abs/2303.01411}{arXiv:2303.01411}). Laws that constrain histories to algorithmically random sequences naturally pair with exchangeable credences encoding inductive symmetries. Using the de Finetti representation theorem, we show that this pairing directly entails the Principal Principle for this framework. We extend the proof to partial exchangeability and provide finite-history bounds that vanish in the infinite limit. The Principal Principle thus emerges as a mathematical consequence of the alignment between nomological constraints and inductive learning. This reveals how algorithmic randomness and exchangeability can illuminate foundational questions about chance, frequency, and rational belief.


\vspace{10pt} 

\end{abstract}


\newpage

\section{Introduction}

We introduce a framework uniting algorithmic randomness with exchangeable credences to address foundational issues in philosophy of probability and philosophy of science. As an application, we show how one might derive the Principal Principle for this framework without circularity.

The Principal Principle (PP)---that rational credence (subjective degree of belief) should match known objective chance---was proposed by Lewis (1980), building on earlier ideas \x{(Miller 1966; Skyrms 1977, 1980)}. Indeed, some form of deference to chance has been implicitly accepted by many who seriously consider the relationship between objective probabilities and subjective beliefs. Lewis articulates this principle: if one knows that an event's chance is $p$, then (absent certain inadmissible information) one’s credence in that event \emph{should} be $p$. For example, if one is certain that a coin’s chance of heads is $0.5$, then (absent knowledge about the outcome of the coin toss) one’s credence that it will land heads should be $0.5$. Lewis called it the \textit{Principal Principle} because it seems to ``capture all we know about chance'' (p. 266).\footnote{Here we use ``chance'' to mean ``objective probability'' or ``nomic probability.'' The precise content of the PP depends on one's account of chance and probabilistic laws. Different metaphysical accounts yield different versions of the PP. Our goal is to derive the PP appropriate for the framework developed here.} 


While widely accepted, the PP raises a fundamental question: why should an objective feature of the world (chance) constrain an agent’s subjective beliefs (credence)? \emph{Prima facie}, chance and probabilistic laws of nature describe how the world behaves, whereas rationality norms prescribe how agents ought to think coherently. Since the PP does not follow from probability axioms alone, rejecting it remains probabilistically coherent. Yet, accepting it as primitive leaves an explanatory gap, especially given that multiple conflicting versions of the PP have been proposed.

Whether the PP can be justified from deeper principles, and if so how, has been a central question in the philosophy of probability, and an unresolved challenge for those who posit chances and probabilistic laws.\footnote{The search for a non-circular proof of the PP has led to some thought-provoking proposals. However, skeptics can and have raised concerns about potential circularity in these approaches. For example, accuracy-based arguments (Pettigrew 2012, 2016) have been criticized for relying on epistemic norms that may already presuppose the deference to chance they seek to justify (Cariani 2017 and Kotzen 2018). In contrast, strategies employing the Humean best-system analysis of chance, such as Schwarz's argument (2014), appeal to a Principle of Indifference, which has well-known difficulties. These difficulties are especially acute for the infinite histories common in modern physics, where choosing a 'uniform' measure on the state space requires substantive assumptions that may themselves depend on the objective chance measure (potentially invoking the PP). Nevertheless, the project in Schwarz (2014) shares conceptual affinities with our own, and his analysis provided inspiration for the approach taken here.} Lewis (1994, p.~484) seems to suggest that only a Humean (e.g. best-system) account of laws could hope to deliver a non-circular justification.

We aim to justify the PP, insofar as we think it can be justified, by broadening the standard toolkit for discussing probability and chance. Specifically, we suggest that a collection of foundational ideas, often treated in isolation, are in fact complementary parts of a unified framework for understanding chance, credence, randomness, relative frequency, and induction. 

The framework has two central components. The first is metaphysical: a general move replacing the standard chance law $L$ with a \emph{statistical constraining law} $L^\star$. This $L^\star$ law excludes deviant-frequency histories and non-random patterns (``maverick worlds'') by mandating that physically possible histories be Martin-L\"of random (Barrett and Chen 2025). \x{Why replace $L$ with $L^\star$ in the first place? They are empirically equivalent for finite observations: they permit the same finite outcome sequences with the same objective probabilities. The choice is therefore theoretical. Barrett and Chen (2025) motivate and develop statistical constraining laws in detail. As summarized in \S3.1, $L^\star$ provides non-Humeans a unified account of deterministic and probabilistic laws as constraints on physical possibility, while allowing Humeans to avoid the Big Bad Bug and the zero-fit problem. The standard $L$-based account remains available. The aim is to articulate an $L^\star$-based alternative so that the two can be compared on their total theoretical merits.}\footnote{\x{We thank an anonymous referee for encouraging us to clarify this.}}

\x{The second component is epistemological: an appeal to inductive assumptions regarding the uniformity of nature, encoded as \emph{exchangeable} credences of rational inquirers, a precise symmetry in their subjective credences. The two components are natural partners. \x{The $L^\star$ law is restrictive enough to guarantee the correct limiting pattern, yet permissive enough to allow every possible finite segment.} This balance makes $L^\star$ laws well suited to exchangeable credences.}

\x{The present paper assembles this framework and develops a further epistemic payoff: combined with exchangeable credences, $L^\star$ yields a non-circular derivation of its corresponding PP.}\footnote{By excluding maverick worlds which are algorithmically atypical, $L^\star$ also provides a nomological grounding for an algorithmic version of Cournot's Principle (for a historical review, see Shafer 2023). \x{We} set this interesting consequence aside for future work.} \x{Our target is thus not the PP for the standard law $L$, but the principle appropriate to the $L^\star$ package. The proof is elementary, relying only on de Finetti's representation theorem, and brings together exchangeability, Martin-L\"of randomness, chance-credence norms, and statistical constraining laws.}

\x{Informally, the proof has three steps. First, de Finetti's theorem represents an agent's exchangeable credences as a mixture over models whose outcomes are independent and identically distributed (iid), with different possible limiting frequencies. Second, $L^\star$ deductively entails the limiting frequency (e.g.\ a limiting frequency of $0.5$ for heads). Hence, accepting $L^\star$ collapses the mixture over models to a single model (e.g.\ the unbiased iid coin-toss model). Third, the agent's resulting single-case credences match the probabilities associated with $L^\star$  and exhibit the statistical properties (namely, independence and resiliency) required by the PP. By contrast, the \emph{corresponding} argument for a standard law $L$ would require a PP-like assumption that a proposition assigned objective probability one must also receive credence one, and would therefore be circular.}


Although we focus on the case of full exchangeability and infinite histories for conceptual clarity, the strategy is robust. It extends to both partial exchangeability and finite histories. In the finite case, one obtains a principle that  approximates the PP with sharp error bounds that provably vanish as the history length approaches infinity. This result strengthens our main proof, showing that the infinite case is the natural and exact limit of the finite principle. 


\x{The framework may also be of interest to subjectivist and pragmatic Bayesians in the tradition of de Finetti and Skyrms. $L^\star$ gives concrete expression to the global patterns to which such agents are committed, and offers a precise alternative for those who find standard notions of objective chance unclear.}\footnote{Moreover, since $L^\star$ guarantees the right global patterns, Bayesian merging of opinions is guaranteed if two Bayesians accept $L^\star$ as the correct law. See Zaffora Blando (2025).}


The framework has additional implications. It solves an open problem about how a statistical constraint law $L^\star$ informs single-case credences. Moreover, the justification for the PP works for both Humean best-system accounts and non-Humean governing conceptions of laws.\footnote{For a related approach, see Mierzewski and Zaffora Blando (2024). Their project explores how the notion of a Humean best description is determined by an agent's inductive assumptions. Motivated by issues raised by Belot (2023, 2024), they relativize the concept of (best-system) chance itself to the agent's credences. \x{Our project instead follows
the more traditional structure of the PP, showing how credences should
defer to a given probabilistic law about the world; the relevant
$L^\star$ law is neutral between Humeanism and non-Humeanism.}} 


The paper proceeds as follows. Section 2 clarifies the PP and its connection to probabilistic laws. Section 3 presents our two key assumptions in the framework. Section 4 provides the proof for full exchangeability. Section 5 extends the proof strategy to partial exchangeability and finite histories. Section 6 discusses additional implications.

\section{\x{The Principal Principle and Probabilistic Laws}}

\x{The Principal Principle (PP) connects two kinds of probability: objective chance, which characterizes a property in the physical world, and subjective credence, which represents an agent's degree of belief. Its basic idea is simple. If an agent knows that the objective chance of heads on a coin toss is $0.5$, and has no information bearing directly on the result, then her credence in heads should also be $0.5$.}

\x{More generally, let $Cr$ be the agent's credence function and let $ch$ be the objective chance function.\footnote{For historical context, one may think of $Cr$ as what Lewis (1980) calls the initial credence function, or sometimes called the ``ur-prior.'' Our framework does not require there be a uniquely correct initial credence. \x{Throughout, $Cr$ is a probability measure on an algebra containing propositions about outcomes, chance hypotheses, and candidate laws. For example, in the approach discussed below, its possibility space may be represented by pairs $\langle\xi,L\rangle$ of a complete history and a candidate law (Chen 2024b). Conditioning on $L^\star$ is ordinary conditionalization, so $Cr(\cdot\mid L^\star)$ is a probability measure over outcome histories concentrated on those compatible with $L^\star$.}} If the agent knows that the chance of an event $A$ is $x$, the PP requires}
\begin{equation}
    Cr(A \mid ch(A)=x)=x.
\end{equation}
\x{The required alignment should also be stable under further information that is irrelevant to whether $A$ occurs. For example, learning the results of earlier coin tosses should not change the agent's credence in the next toss when those results provide no information about it. Following Skyrms (1977, 1980), we call this stability \emph{resiliency}. If $K$ is admissible with respect to $A$---that is, if it does not bear directly on whether $A$ occurs---then}
\begin{equation}
    Cr(A \mid ch(A)=x \wedge K)=x.
\end{equation}
We assume throughout that the relevant conditional credences are defined: any conditional credence $Cr( \cdot \mid B)$ is understood to be asserted only when $Cr(B)>0$. 
 This is a version of what Lewis (1994) calls the \emph{Old Principle}.

\x{This is the traditional formulation of the PP, but it is not always the most natural formulation for physical theories with probabilistic postulates. Theories such as the Mentaculus (Loewer 2020), the Wentaculus (Chen 2024a), and Bohmian mechanics (Goldstein 2025) do not begin by assigning primitive single-case chances to every event. Instead, a general probabilistic law $L$ includes an objective probability measure over possible initial microstates. Together with the dynamics, this induces a probability measure $\mu_L$ over complete microscopic histories, which are fully specified ways for the world to unfold. Probabilities for subsequent macroscopic events---such as formation of stars, melting of ice cubes, and results of coin tosses---are then derived from this global measure (Lewis 1994, pp.~477--478; Loewer 2004). We call $\mu_L$ the law's associated \emph{nomic measure}. The PP can therefore be reformulated as a direct connection between subjective credence $Cr$ and this measure $\mu_L$, with $\mu_L$ playing the role of chance:}

\begin{description}
\item[Law-to-Credence Principle (LCP)] For any event $A$,
\begin{equation}
    Cr(A \mid L) = \mu_L(A).
\end{equation}
\item[Resiliency] For any $K$ admissible with respect to $A$ (relative to $L$),
\begin{equation}
    Cr(A \mid L \wedge K) = \mu_L(A).
\end{equation}
\end{description}
\x{LCP says that, conditional on $L$, the agent's credence in an event should equal the probability assigned by the law's nomic measure. Resiliency says that this alignment persists after conditioning on further admissible information.} We understand the PP as the conjunction of LCP and Resiliency.

For example, consider an infinite $\omega$-sequence of coin tosses $\langle r_1, r_2, \ldots \rangle$ governed by law $L$: 
\begin{description}
    \item[$L$:] Each element in the $\omega$-sequence of coin tosses $\langle r_1, r_2, \ldots \rangle$ is determined independently and with an unbiased probability of heads and tails.
\end{description}
One might think of $L$ as describing the fundamentally random process where one starts with a sequence of spin-$1/2$ particles each in an eigenstate of $z$-spin, then measures their $x$-spins in turn.

This law $L$ uniquely determines a nomic measure $\mu_L = \beta_{0.5}$ (the unbiased iid Bernoulli product measure on the binary outcome space).  LCP requires that, for any toss $j$:
\begin{equation}
        Cr(X_j=H \,|\, L)=\mu_L(X_j=H)=\beta_{0.5}(X_j=H)=0.5.
\end{equation}
The agent's credence should align with the known nomic measure. Resiliency ensures this remains true even after observing any finite history of previous tosses:
\begin{equation}
        Cr(X_j=H \,|\, L \wedge K)=\mu_L(X_j=H)=\beta_{0.5}(X_j=H)=0.5,
\end{equation}
where $K$ encodes past results. In other words, the agent's credence given $L$ is probabilistically independent of $K$. Throughout the paper, we use this coin-tossing example with $\beta_{0.5}$ for concreteness, though our framework generalizes to other probabilistic structures. 

\x{The LCP also illustrates why the PP matters in empirical confirmation. For concreteness, we focus on Bayesian confirmation theory, though the point generalizes. Bayesian confirmation of rival probabilistic laws requires evidential likelihoods: the agent's conditional credences in the evidence, given each law. For example, if $L$ and $L'$ are two rival probabilistic laws, and $\mu_L$ and $\mu_{L'}$ assign objective probabilities $0.2$ and $0.7$ to event $E$, comparing $L$ against $L'$ requires the agent to take $Cr(E \mid L) = 0.2$ and $Cr(E \mid L') = 0.7$, which is an application of the LCP. Without this coordination between objective and subjective probabilities, the evidential likelihoods are underdetermined, and Bayes factors cannot adjudicate between $L$ and $L'$ in a principled way (Strevens 2017, pp.\ 31--36). The PP is thus presupposed by ordinary statistical practice. In our framework, the required coordination is a theorem rather than a posit.}

\section{The Framework}

We now describe the two key components of the framework: a statistical constraining law $L^\star$ excluding deviant patterns, and exchangeable credences encoding inductive assumptions. 

\subsection{\x{Constraining Law $L^\star$}}

Barrett and Chen (2025) introduce a \textit{statistical constraining law} $L^\star$ that is in an important sense stronger than an ordinary probabilistic law.  Where a standard probabilistic law $L$ assigns measure~1 to certain patterns while tolerating measure-zero exceptions,  a statistical constraining law $L^\star$ makes those patterns nomologically necessary by requiring sequences to be algorithmically random with respect to a specified reference measure. 

Consider a concrete example of such a statistical constraint law: 
\begin{description}
    \item[$L^\star_{\beta_{0.5}}$:] The $\omega$-sequence of coin tosses $\langle r_1, r_2, \ldots \rangle$ is \emph{Martin-L\"of random} with unbiased frequencies of heads and tails, i.e. with respect to the Bernoulli iid measure $\beta_{0.5}$.
\end{description}
For notational simplicity,  we use $L^\star$ to refer to $L^\star_{\beta_{0.5}}$ throughout this paper, unless noted otherwise. 

This law does two things. First, it guarantees that heads occurs exactly 50\% of the time in the infinite limit---not merely ``almost surely.'' Second, it excludes all algorithmically detectable patterns, ruling out sequences such as the alternating heads-tails sequence $(1010...)$ that fail Martin-L\"of randomness tests.\footnote{See Li and Vitányi (2019), Dasgupta (2011), Martin-Löf (1966) for introductions to algorithmic randomness; Eagle (2021) for philosophical issues; Hájek (2023) on interpretations of probability.} 

Crucially, Martin-L\"of randomness is a tail property---it concerns the behavior of the entire infinite sequence, not any finite initial segment. In fact, any specific finite string (say, a million heads in a row) can appear at the beginning of a Martin-L\"of random sequence. What matters for randomness is whether the infinite continuation exhibits algorithmic patterns. This tail property is essential for our framework: while $L^\star$  constrains the infinite sequence to have limiting \x{frequency} $0.5$ and no algorithmic patterns, it places no restrictions on finite initial segments. \x{As we explain in \S3.3, this does not render finite observations evidentially idle. Together with exchangeability, $L^\star$ supplies the subjective likelihoods needed for Bayesian confirmation.}

The framework generalizes naturally to biased cases: for example, $L^\star_{\beta_{0.9}}$ allows only $\omega$-sequences with 90\% heads and 10\% tails, and only those that satisfy the associated algorithmic tests with respect to $\beta_{0.9}$. In fact, this generalizes beyond iid measures to any computable measure.\footnote{Martin-L\"of himself showed how to provide an algorithmic notion corresponding to an arbitrary computable probability distribution (1966, 612--4). See Porter (2020) for a recent survey on biased algorithmic randomness. 
We consider Bernoulli measures and, in section~5.1, a stationary Markov measure as examples. More generally, if $\mu$ is a computable stationary ergodic measure, then $\mu$-Martin-L\"of randomness guarantees the corresponding limiting relative frequencies. More generally still, for an arbitrary computable measure $\mu$, Martin-L\"of randomness \textit{guarantees} the satisfaction of its effective almost-sure statistical laws whatever they may be (Zaffora Blando, 2025). In this most general case, the statistical laws need not include convergent relative frequencies.
The framework can further be extended to a non-computable measure $\mu$ by using randomness relative to a Turing machine with an oracle for $\mu$. The issues here are subtle, which we leave to future work. For some recent discussions of the connections of computability, chance, and credence, see Belot (2023, 2024).} 

\x{Since no maverick worlds are nomologically possible under $L^\star$, an
agent who accepts $L^\star$ must assign credence $1$ to the $0.5$
limiting frequency by logical consistency and probabilistic
coherence, a feature critical for our derivation in \S4.}

It is worth clarifying that this approach is not simple frequentism. Frequentism  seeks to define probability in terms of frequencies of events. By contrast, the law $L^\star$ starts with a theoretical measure (here $\beta_{0.5}$) that is part of the law's content. This measure, over the entire space of possible histories, is not derived from frequencies but rather serves as the standard against which sequences are judged to be algorithmically random. For example, the same history can be judged random with respect to $\beta_{0.5}$ but non-random with respect to $\beta_{0.9}$. Moreover, algorithmic randomness goes beyond what simple frequentism can capture. Historically, von Mises's account of frequentism (extended by Church and Wald) was shown to be inadequate to rule out nonrandom patterns such as Ville's sequence, while Martin-L\"of randomness succeeds in excluding them. It remains an open question whether Martin-L\"of randomness can be fully captured in frequentist terms (see Barrett and Chen (2025, \S6) and Dasgupta (2011) for more discussions). Nevertheless, $L^\star$ is compatible with, though does not entail, sophisticated versions of frequentism, such as the Humean best-system account of chance.

In our framework, we systematically replace standard probabilistic laws $L$ with these statistical constraining laws $L^\star$. 

Two considerations motivate using $L^\star$ laws in our framework. The first is clarity about the objective side of the PP---what exactly should rational credence defer to? Following Lewis (1994, pp.\ 477--478), we treat single-case chances as derived from probabilistic laws carrying nomic measures, \x{rather than as primitive chance propositions}. The law $L^\star$ makes this transparent: it specifies a unique reference measure ($\beta_{0.5}$) and restricts physical possibility to sequences that are algorithmically random with respect to that measure. This gives a clear target for the LCP and Resiliency requirements, \x{and a fully precise propositional content for the object of conditioning}.

\x{Secondly, $L^\star$ has attractive theoretical virtues (Barrett and Chen 2025). Though empirically equivalent to $L$ for finite observations (\S1), it resolves longstanding puzzles on both sides of the Humean divide. For non-Humean governing accounts, such as minimal primitivism, $L^\star$ unifies how laws govern (Chen and Goldstein 2022, Chen 2024b). A standard law $L$ rules out no history, and interpreting the numbers it assigns to propositions often suggests gradable primitives such as propensities. In contrast, $L^\star$  governs by constraining; it rules out the maverick worlds as physically impossible, just as $F=ma$ rules out its non-solutions. Hence, a single primitive relation, constraining physical possibility, covers deterministic and probabilistic laws alike. For Humean best-system accounts, two familiar problems are dissolved. The Big Bad Bug never arises, since the maverick worlds that generate the contradiction are nomologically impossible. There is thus no need for the revised, arguably less intuitive, versions of the PP proposed in response (Lewis 1994). The zero-fit problem is also eliminated. Given an infinite outcome sequence, there is much underdetermination among standard $L$ laws, and Humeans would add \emph{fit} to choose the winning best system. However, on each infinite sequence, the fit is exactly zero given each law, so it cannot be the tie breaker (Elga 2004). In contrast, $L^\star$ laws need no criterion of fit, since at most one candidate $L^\star_{\beta_\theta}$ is true of the actual sequence, so best systems can be compared by simplicity and informativeness alone, as with non-probabilistic laws (Barrett and Chen 2025, \S5).}


In this framework, the PP for $L^\star$ takes on the following forms of  LCP and Resiliency: 
\begin{equation}\label{LCPstar}
        Cr(A \,|\, L^\star)=\mu_{L^\star}(A)=\beta_{0.5}(A).
\end{equation}
For any admissible $K$ (with some to-be-defined notion of admissibility):
\begin{equation}\label{Resiliencystar}
  Cr(A \,|\, L^\star \wedge K)=\mu_{L^\star}(A)=\beta_{0.5}(A).
\end{equation}

We turn next to the role of exchangeability in the framework.

\subsection{\x{Exchangeability}}

Exchangeability expresses a symmetry property of probability measures. Applied to credences, it describes how an agent's credences treat outcomes in a sequence of trials (de Finetti, 1937, 1974, 1975). Concretely, if one's credences are exchangeable, the subjective probability one assigns to seeing a particular number of heads and tails in $n$ tosses of a coin, fair or biased, depends only on the number of heads and tails, not the order in which they occurred. 

An agent's credences are exchangeable over a sequence of $n$ tosses if and only if her credence for each reordering of the sequence is the same. And her credences are exchangeable over an infinite sequence of results if and only if they are exchangeable over every finite initial segment. We say that her credences are exchangeable \emph{simpliciter} when they are exchangeable over an infinite sequence.  Formally, $Cr$ is exchangeable if and only if for every $n\geq1$, every sequence of outcomes $(x_1, x_2, ..., x_n)$, and every permutation $\pi$ of the indices $\{1,2,...n\}$, the following equality holds:
\begin{equation}\label{exchangeability}
    Cr(X_1=x_1, X_2=x_2, ..., X_n=x_n)=  Cr(X_{\pi(1)}=x_1, X_{\pi(2)}=x_2, ..., X_{\pi(n)}=x_n).
\end{equation}

In what follows we consider exchangeability of credences, which is a property of subjective credences, not objective chance. It does not require that the objective chance of ``heads on toss 3'' is in fact equal to the objective chance of ``heads on toss 7'', i.e. $ch(X_{3}=H)=ch(X_{7}=H)$---rather, it just means that the inquirer with exchangeable credences sees no reason to distinguish one position in the sequence from another \textit{given her present evidence}. Credences that begin as exchangeable may fail to be exchangeable after an inquirer conditions on new evidence if that evidence suggests that order does matter.

The exchangeability of one's credences can be thought of as a revisable \textit{inductive assumption}: in the absence of specific information about different times or trials, if an agent's credences are exchangeable, she is committed to treat those trials as relevantly the same (Diaconis and Skyrms, 2018). Exchangeability sets the stage for de Finetti's representation theorem, which states that if one's credences are exchangeable over a particular set of outcomes, then one can think of the process generating those outcomes as being a mixture over iid models. We will consider this theorem in some detail later. Let us further illustrate the connections between exchangeability and induction. 



First, an exchangeable credence is symmetric for all trials, since permutation invariance implies that $\forall i\neq j, \, Cr(X_i=H)=Cr(X_j=H)$.\footnote{Proof: By exchangeability, we have $Cr(X_i=H, X_j=T)=Cr(X_j=H, X_i=T)$. Hence, $Cr(X_i=H)=Cr(X_i=H, X_j=H)+Cr(X_i=H, X_j=T)=Cr(X_i=H, X_j=H)+Cr(X_j=H, X_i=T)= Cr(X_j=H)$.} Hence, credence exhibits temporal stability---it cannot have time-dependent biases such as $Cr(X_{500}=H)>Cr(X_{50}=H)>Cr(X_{5}=H)$. This symmetry expresses a form of temporal uniformity, though not necessarily independence across time. 

Moreover, as we shall see, the de Finetti representation theorem deepens these connections. Since an exchangeable credence can be represented as a mixture of iid processes with unknown bias $\theta$, unless one is certain about $\theta$, observing outcomes updates beliefs about $\theta$. Specifically, if the prior on $\theta$ is non-dogmatic, observing  $X_{1}=H$ will update the credence mixture to one more peaked around higher values of $\theta$, yielding  $Cr(X_{2}=H | X_{1}=H)\geq Cr(X_{2}=H)$. This means past experiences are inductively relevant to beliefs about the future, with learning occurring through Bayesian updating on $\theta$ in light of observed frequencies. This expresses a probabilistic commitment that the future will resemble the past. 

Furthermore, under the mixture representation given by de Finetti's theorem, the agent assigns probability $1$ to the existence of a limiting relative frequency, because each component iid model assigns probability $1$ to its own limiting frequency $\theta$. Thus an agent with exchangeable credence is $100\%$ confident that there exists a long-run statistical pattern in the sequence. Exchangeable credence thereby commits one to inductive learning and to confidence in the existence of convergent, learnable patterns. Importantly, while exchangeability itself only guarantees identical marginals, it is conditional on $\theta$ that the $X_i$'s are independent and identically distributed. (See Skyrms 2014 and Zabell 2005 for more discussions.) 


Because of their tight connections, the epistemic status of exchangeable credences mirrors that of induction itself. As Hume (1748) effectively argued, inductive reasoning has no non-circular epistemic justification. Many treat induction as properly basic---a foundational principle of rational inquiry that needs no further grounding (for a survey, see Henderson 2024). Without induction, empirical knowledge and scientific progress would be impossible. If we accept this view, then exchangeable credences inherit similar epistemic standing. If we hold exchangeable credences over coin tosses, we are making an inductive judgment that position in the sequence does not matter absent specific evidence to the contrary. This mirrors scientific practice: researchers routinely assume that experimental trials are exchangeable (that Monday's electrons behave like Friday's electrons). Indeed, such methodological assumptions are constitutive of empirical scientific inquiry as we find it. Insofar as we are entitled to adopt standard inductive methods without special justification, exchangeability---being a precise formalization of several of our inductive commitments in many contexts---is similarly warranted. Put another way, exchangeability holds precisely when we believe that the probability of outcomes does not depend on their order, and our inductive practices indicate we very often believe precisely this. This makes exchangeability a principled stance grounded in the broader framework of inductive reasoning.\footnote{\x{This commitment can be understood through two Bayesian lenses (see Lin 2024 for an overview). For subjective Bayesians, exchangeability is a permissible and often well-motivated property of the agent's \textit{de facto} beliefs about the world; the PP then follows for any such agent who also accepts $L^\star$. For objective Bayesians who treat induction as a requirement of rationality, exchangeability may be a warranted normative constraint, and the PP becomes a derivative requirement given $L^\star$.}}


That said, our framework does not assume that an agent's credences are exchangeable \textit{tout court}. What we need is conditional exchangeability: the agent's credences over outcome strings are exchangeable \emph{given} a $\star$-law that characterizes the global properties of the sequence. Formally, $Cr(\cdot \mid L^{\star})$ is exchangeable. As we explain in \S3.3 and \S4.3, the randomness built into $L^\star$ harmonizes with exchangeable credences. 

Importantly, conditional exchangeability is compatible with failure of unconditional exchangeability. As we note in \S5.1, a Bayesian inquirer's unconditional credences are typically an epistemic mixture over different uniformity patterns, including both iid and non-iid models; such mixtures generally break exchangeability at the unconditional level---even though, after conditioning on a particular $L^\star$ law, the agent's conditional credences may be exchangeable.  Our argument in \S4 requires only the conditional exchangeability assumption.




\subsection{Harmony}

The two components of the framework---statistical constraining laws and exchangeable credences---are natural partners. However, they are conceptually independent. An agent could coherently adopt exchangeable credences while believing in a standard probabilistic law $L$ (which permits maverick worlds). Conversely, an agent could believe in a constraining law $L^\star$ while holding non-exchangeable credences. 

Their harmony, rather, is one of conceptual resonance, revealed by a key consequence of de Finetti's theorem and Martin-L\"of randomness: an agent with exchangeable credences assigns credence $1$ that the sequence of events will exhibit \emph{some} limiting frequency $\theta$ and be Martin-L\"of random with respect to the corresponding measure, $\beta_\theta$. 

While an exchangeable prior does not commit the agent to any particular value of $\theta$, it commits the agent, with credence~1, to the belief that the world exhibits the pattern dictated by \emph{some} statistical constraint law. The agent's credence is concentrated on the set of worlds that are nomologically possible according to \emph{some} $L^\star_{\beta_\theta}$. An agent who adopts exchangeability is therefore already epistemically aligned with the form of lawfulness that $L^\star$ makes precise. Accepting a specific law, such as $L^\star_{\beta_{0.5}}$, then simply collapses this epistemic mixture onto a single, determinate nomological possibility.

\x{This} \x{also explains why the tail character of $L^\star$ does not make finite observations evidentially idle. As we show in \S4, conditional exchangeability given each candidate law $L^\star_{\beta_\theta}$ yields $Cr(E \mid L^\star_{\beta_\theta}) = \beta_\theta(E)$ for any finite evidence string $E$. These subjective likelihoods support Bayesian updating over candidate laws, even though no finite string conclusively rules out any candidate. This is precisely the coordination between objective and subjective probabilities that principled comparison of rival chance hypotheses requires (\S2). For standard probabilistic laws, the coordination is often posited, via the PP itself or an equivalent principle, rather than derived. For $L^\star_{\beta_\theta}$, we show that it can be derived non-circularly.}

\section{Proof of the Principal Principle}

We now show how the framework of exchangeable credences and the constraining law $L^\star$ yields the PP. 

\subsection{The Proof}

We will start by deriving the PP for an agent who is fully committed to the law $L^\star=L^\star_{\beta_{0.5}}$ and whose conditional credences are exchangeable.  
For this particular  $\star$-law,  the nomic measure is the unbiased iid Bernoulli measure, $\beta_{0.5}$, and the relative frequency of the two-valued sequence of heads and tails has a limiting relative frequency of $0.5$ heads. As noted, the PP for $L^\star$ is the conjunction of LCP \eqref{LCPstar} and Resiliency \eqref{Resiliencystar}.

\begin{theorem}[Single-case credence]

\noindent
An agent with exchangeable credences who fully believes that $L^\star$ holds is committed to treating the sequence of outcomes as an unbiased Bernoulli iid process. Her conditional credence measure is therefore identical to the nomic measure. In other words, if $Cr(\cdot\mid L^\star)$ is exchangeable, then $Cr(\cdot\mid L^\star)=\beta_{0.5}(\cdot)$.
\end{theorem}

\begin{proof}
The proof is a direct consequence of de Finetti's representation theorem. As presented by Diaconis and Skyrms (2018), for any exchangeable and extendable probability assignment $P$ to sequences of two values, there exists a unique prior probability $\mu$ on $[0, 1]$ such
that for any finite binary sequence $e_1, e_2, \ldots , e_n,$
\begin{equation}\label{integral}
P(e_1, e_2, \ldots , e_n) = \int_{0}^{1} \theta^s(1-\theta)^{n-s} \mu(d\theta),
\end{equation}
with $s=e_1 + \cdots + e_n$, the number of heads among ${e_1, e_2, . . . , e_n}$. Moreover $P$ assigns probability one to the event
\begin{equation}
\mbox{\{proportion of heads in the first $n$ places has a limit $\theta$\}} 
\end{equation}
and
\begin{equation}
P(\theta \leq x)=\mu(0, x].
\end{equation}

We specialize the function $P$ to an agent's conditional credence, $Cr_{L^\star}=Cr(\cdot \mid L^\star)$, which obeys the probability axioms and is exchangeable. (Since $Cr_{L^\star}$ is exchangeable over all finite sequences, it is exchangeable and extendable.) Informally, the second part of the theorem means that a Bayesian with an exchangeable credence $Cr_{L^\star}$ is sure that long-run frequencies exist, the first part means that $Cr_{L^\star}$ may be represented as a mixture of iid processes (a mixture of $\beta_{\theta}$ distributions), and the third part means that the mixing distribution $\mu$ is uniquely identified as the long-run frequency distribution \x{of $Cr_{L^\star}$} (Diaconis and Skyrms 2018, p.137).

The crucial step is this: if an agent fully believes that $L^\star$ holds,  she must fully believe its logical consequences. Since $L^\star$ logically entails that the limiting frequency of heads is $\theta=0.5$, her credence in that proposition must be $1$:
\begin{equation}
    Cr(\theta=0.5 | L^\star) = 1.
\end{equation}
This collapses the mixture in the de Finetti representation. The weighting measure $\mu$ becomes concentrated at $\theta=0.5$. In other words, if her credences are exchangeable on the sequence, she believes that the outcomes in the sequence can be represented as unbiased iid coin tosses (with $\theta=0.5$). For any finite sequence $e_1, e_2, \ldots , e_n,$ the integral (\ref{integral}) thus becomes: 
\begin{equation}
Cr(e_1, e_2, \ldots , e_n | \,L^\star) = 0.5^s (1-0.5)^{n-s}=0.5^n = \beta_{0.5}(e_1, e_2, \ldots , e_n).
\end{equation}
This equation establishes that the agent's conditional credence for any fine-grained history is identical to the probability assigned by the unbiased Bernoulli iid measure, $\beta_{0.5}$. Because this holds for all finite sequences, it follows that the agent's credences about all events in the sequence are also governed by the $\beta_{0.5}$ measure and are probabilistically independent. 
\end{proof}

A similar theorem holds for biased Martin-L\"of random sequences. The only difference is the $\star$-law and the limiting relative frequencies that the agent fully believes. One simply replaces $\beta_{0.5}$ with the relevant biased measure $\beta_\theta$. Using the general version of de Finetti representation theorem (Diaconis and Skyrms, 2018 p.138), the same reasoning applies to multi-valued outcome spaces, with the mixing measure concentrated at the limiting relative frequencies specified by the appropriate $\star$-laws. We discuss other generalizations in \S5. 


The \textit{single-case credence} theorem immediately \x{yields LCP, since $Cr(\cdot\mid L^\star)=\beta_{0.5}(\cdot)$.}

Furthermore, since the agent's conditional credence measure is an iid measure, her beliefs about distinct trials are probabilistically independent. This guarantees a precise version of Resiliency. Her credences remain invariant when conditioning on any admissible evidence $K$---that is, any information that is not itself directly informative about the outcome in question. More precisely, $K$ is admissible with respect to event $A$ and law $L^\star$ if and only if $K$ is screened off from $A$ under the nomic measure $\beta_{0.5}$ (i.e. $\beta_{0.5}(A\mid K)=\beta_{0.5}(A)$) and $\beta_{0.5}(K)>0$. For example, the following holds: 
\begin{equation}
Cr(e_{100}, \ldots , e_{200} | \,L^\star) = Cr(e_{100}, \ldots , e_{200} | \,L^\star \, \wedge e_{1}, \ldots , e_{99} ) =Cr(e_{100}, \ldots , e_{200} | \,L^\star \, \wedge e_{300} ) = 0.5^{101}.
\end{equation}
In this way, her credences are resilient. 





\subsection{Avoiding Circularity}

In contrast, a circular argument for the PP proceeds as follows for a standard probabilistic law $L$: $L$ assigns nomic measure~1 to the set of histories with the correct limiting frequency $\theta=0.5$. \x{``Almost certain'' is thus the strongest qualification the law itself supplies. Notice that $\mu_L(\theta=0.5)=1$ is a fact about the law's measure, not yet a claim about rational credence.} If one then assumes a bridging principle---that an event with nomic measure~1 warrants a rational credence of $1$---an agent with exchangeable credences becomes confident with credence $1$ that $\theta=0.5$. The de Finetti representation then ensures her credences align with the law. Such an argument is circular because the bridging premise (``nomic measure 1 implies rational credence 1'') is an instance of the very principle it aims to derive. Other attempts to justify the PP grapple with this circularity problem in different ways (see footnote 2). 

Our derivation avoids this circularity. $L^\star$ does not merely assign nomic measure 1 to the correct frequency $\theta=0.5$; it \emph{logically entails} it. 
In contrast, an agent who accepts $L^\star$ as the correct law must, on pain of probabilistic incoherence, have credence~1 that $\theta=0.5$. This conclusion follows from logical entailment, not from a deference principle. De Finetti representation theorem then ensures that the agent’s exchangeable credence aligns with the measure specified by the law, so the single-case credence matches the stated chance. Thus PP holds as a direct consequence of the law's content together with the agent’s inductive commitments.

In order to see that neither assumption---the law $L^\star$ \x{or exchangeable credences}---covertly presupposes the PP, note that neither entails the principle on its own. The law $L^\star$ is a metaphysical posit about the scope of nomological possibility. It is compatible with any probabilistically coherent prior, including non-exchangeable ones. Exchangeability formalizes an inductive stance but is agnostic about objective chances; indeed, de Finetti famously used it to argue that objective chance is a superfluous concept (1974-75). Exchangeable credences, rather, just allow the agent to treat the outcomes of tosses \textit{as if} they were the result of an iid process.  Our argument appeals to features of \textit{both} the inquirer's exchangeable credences and her commitment to $L^\star$.

\subsection{The Role of Algorithmic Randomness}

While the proof of the PP primarily relies on the frequency constraint imposed by the $L^\star$ law, the consistency of the agent's commitments crucially depends on the randomness property provided by Martin-Löf randomness. This is related to the conceptual harmony discussed in \S3.3. 

In particular, because Martin-Löf randomness concerns the tail property of the sequence, it allows the assumption of full exchangeability of the credence (conditional on $L^\star$) by ensuring that any finite segment of outcomes can and will appear somewhere in the infinite sequence. Recall that exchangeability requires equality (\ref{exchangeability}) to hold for every $n\geq1$, every sequence of outcomes $(x_1, x_2, ..., x_n)$, and every permutation $\pi$ of the indices $\{1,2,...n\}$. With an arbitrary statistical constraint, one would worry regarding the epistemic permissibility of adopting fully exchangeable credences. With Martin-L\"of randomness, however, the logical guarantee that sufficiently random sequences include all finite patterns ensures room for full exchangeability as an assumption about epistemic credences.


This contrasts with some frequentist constraints that treat certain finite strings as ``insufficiently random'' and effectively exclude them. As Schwarz (2014) points out, such exclusions preclude fully exchangeable credences.  $L^\star$ law avoids this problem: its permissiveness over finite patterns makes it probabilistically coherent for an agent who accepts $L^\star$  to adopt conditionally exchangeable credences.

\x{Our strategy of deriving the PP generalizes beyond Martin-L\"of randomness. The proof of Theorem~1 uses only that the $\star$-law logically entails the relevant limiting frequencies. To play the same broader role in our framework, an alternative randomness notion should also characterize a measure-one class relative to the reference measure, exclude the relevant effectively detectable patterns, and impose no restrictions on finite initial segments, which ensures compatibility with conditional exchangeability. For example, Schnorr randomness has these features too (see for example Nies 2009, Theorems~3.5.19 and 3.5.21) and can play the same role in the derivation. We use Martin-L\"of randomness to develop a concrete version of the framework. Other suitable randomness notions (see Porter 2016, 2021; Barrett and Chen 2025 \S3) would provide corresponding $\star$-law frameworks.\footnote{\x{Thanks to an anonymous referee for encouraging us to clarify this point.}} Section~5 develops further generalizations along other dimensions.}

\section{Generalizations}

We proved the PP in a core case---assuming full exchangeability and an $L^\star$ law for infinite histories---to prioritize the conceptual issues. While this case is instructive in its simplicity, the proof strategy is robust and extends to more complex inductive commitments and to finite histories. 

\subsection{Partial Exchangeability}

Our main proof (\S4.1) requires only that the agent's credence is exchangeable \emph{conditional} on the law $L^\star$. As mentioned in \S3.2, an agent's unconditional prior, however, may be more complex. A rational agent might begin an investigation with a general commitment to uniformity but be uncertain about the specific pattern of dependence. Her unconditional credence, $Cr(\cdot)$, could therefore be a mixture of different symmetry structures. 


Partial exchangeability generalizes the iid-style symmetry (complete permutation invariance) to accommodate such cases. As a key example, consider stationary Markov exchangeability. A probability function is Markov exchangeable if and only if the probability of any finite sequence depends just on its initial state and the number of times each possible transition occurs. This means that if two different sequences of events start with the same event (say, H) and have the exact same ``transition statistics'' (the same numbers of H-to-H, H-to-T, T-to-H, and T-to-T transitions, summarized in a matrix $\mathbf{P}$), then they must be assigned the same probability. 
Moreover, such a probability function is \textit{stationary} if and only if its statistical properties do not change over time.  Freedman (1962) proves a de Finetti-style representation theorem, showing such a probability function can be uniquely represented as the integral: 
\begin{equation}
        Cr(X)=\int_{S} \gamma_\mathbf{P} (X) d\mu_S(\mathbf{P}),
\end{equation}
where $\mathbf{P}$ is a Markov transition matrix in the space $S$ of possible transition matrices, the stationary Markov measure $\gamma_\mathbf{P}$ induced by $\mathbf{P}$, and $\mu_S$ a probability measure on that space. Importantly, such a probability function also has convergent patterns about limiting transition statistics, which (except in the degenerate case where a Markov exchangeable model is also an exchangeable one) will be different from those entailed by a fully exchangeable probability function. For generalizations to Markov exchangeability, see Diaconis and Freedman (1980a).



Consider an agent who is committed to a general form of inductive assumption about the uniformity of nature but unsure about the specific form of the pattern. For simplicity, suppose her credence is a mixture of an exchangeable model and a stationary Markov exchangeable model: 
\begin{equation}
        Cr(X)= w_I \int_0^1 \beta_\theta(X)d\mu_I(\theta) + w_M\int_{S} \gamma_\mathbf{P} (X) d\mu_S(\mathbf{P}),
\end{equation}
where $w_I$ and $w_M$ are the prior weights on the two models. What happens when this agent learns and conditionalizes on a specific constraining law? 

\textbf{Case 1:} The law is an iid-type constraining law $L^\star$. Suppose the agent learns that the law is $L^\star$, which nomologically requires an iid pattern with a $0.5$ frequency. This law is incompatible with the dependent transition statistics predicted by any stationary Markov model (except in the degenerate case).  Consequently, the second term in the agent's prior has a posterior weight of $0$. Her conditional credence collapses to the first term. The proof in \S4.1 shows that the integral further collapses to the Bernoulli iid measure associated with $L^\star$, yielding the standard PP:
\begin{equation}
        Cr(X \,|\,L^\star)=   \beta_{0.5}(X).
\end{equation}

\textbf{Case 2:} The law is a Markov-type constraining law $M^\star$. Now, suppose the agent learns that the law is  $M^\star$, which nomologically requires a specific stationary Markovian pattern of transitions described by the matrix $\mathbf{P_M}$ stipulated by $M^\star$.  On $M^\star$, the $\omega$-sequence of events is Martin-L\"of random with respect to the nomic measure $\gamma_{\mathbf{P_M}}$ for this $\mathbf{P_M}$. The law $M^\star$ is incompatible with the iid frequency patterns predicted by the fully exchangeable model. Consequently, the first term in the agent's prior has a posterior weight of $0$. Her conditional credence collapses to the second term. By a proof similar to that in \S4.1, the integral further collapses to $\gamma_{\mathbf{P_M}}$, yielding a Markovian version of the PP:
\begin{equation}
        Cr(X \,|\,M^\star)= \gamma_{\mathbf{P_M}} (X).  
\end{equation} 
Such a credence is as resilient as allowed by the memoryless property of a stationary Markov measure: future credences depend only on the current state, not on earlier states. 

More generally, we can consider unconditional credences that are spread out over additional exchangeability models with different symmetries. Analogous versions of the PP will follow by conditioning on the appropriate $\star$-law.

\subsection{Finite Histories}

We started with an $L^\star$ law for infinite histories. Is this reliance on infinite histories problematic?

\x{Insofar} as we understand physical systems as evolving indefinitely, under classical or quantum dynamics, any finite probabilistic process can be understood as a segment of an infinite history unfolding according to a general law such as $L^\star$. While a specific object like a coin cannot be tossed forever (cf. H\'ajek 2009), the underlying sequence of elementary events in our universe can be, and typically is, represented as unbounded in many cosmological theories.

Such considerations, however, do not preclude interest in the finite case. One might still want to analyze it for other reasons, such as to explore how a similar principle might be proven under more restrictive conditions. In this spirit, we show here how finite exchangeability when paired with a finite version of an $L^\star$ law entails a principle that approximates the PP with precise error bounds, which is related to the New Principle (cf. Hall 1994, Thau 1994, Lewis 1994). We leave their comparison to future work. 

The first step is to define a finite version of $L^\star$ law. Since there is no canonical definition of algorithmic randomness for finite strings, we need to make some conventional choices.\footnote{Another option is to consider a vague finite $L^\star$ law in the sense of Chen (2022).} 
For concreteness, suppose the universe has a temporal length of $W$ trials. We shall use $\beta_{0.5}$ as the reference measure, and define a law $L^\star_{\epsilon, z}$ for binary strings of length $W$ as the conjunction of the following constraints:
\begin{enumerate}
    \item Frequency constraint ($Freq_{\epsilon}$): the relative frequency of heads is in the range $(0.5-\epsilon, 0.5+\epsilon)$. For definiteness, let $\epsilon = \sqrt{\log W/W}$. If, say, $W=2^{100}$, we have $\epsilon \approx 2^{-47}$. \x{Logarithms are base 2 throughout.}
    \item Randomness constraint ($Rand_z$): \x{Let $h(s)$ be the number of heads in string $s$. There are $\binom{W}{h(s)}$ length-$W$ strings with $h(s)$ heads; call this set the \emph{frequency class} of $s$. Almost all of its members (in cardinality) have nearly maximal \emph{conditional prefix Kolmogorov complexity} given the class (Li and Vit\'anyi 2019, p.~206), about $\log\binom{W}{h(s)}$. We require $s$ to be a typical member of this class, with $K_M(s\mid W,h(s)) \geq \log\binom{W}{h(s)}-z$, where $M$ is a fixed universal prefix-free Turing machine and $z$ is the permitted randomness deficiency (small compared to $W$, allowing some small degree of pattern). In other words, the ordering in $s$ may be at most $z$ bits more compressible than a typical ordering with the same number of heads. This instantiates the standard notion of randomness deficiency relative to a finite set (Li and Vit\'anyi 2019, pp.~410--412).} For definiteness, let $z=\sqrt W$. When $W=2^{100}$, $z=2^{50}$.  
\end{enumerate}
In symbols, $L^\star_{\epsilon,z}=Freq_{\epsilon}\cap Rand_z$. \x{Let $\neg Rand_{z}$ denote the sets of length-$W$ strings that violate $Rand_{z}$.}


The following four-step proof then shows that an agent starting with an
exchangeable credence \x{$Cr$} will, after conditioning on $L^\star_{\epsilon, z}$, have a new credence $Cr_{L^\star_{\epsilon, z}}$ that is very close to an unbiased iid probability distribution.

\textbf{Step 1.} Conditioning the agent's exchangeable credence $Cr$ on $L^\star_{\epsilon, z}$ \x{can break} exchangeability. This is because $L^\star_{\epsilon, z}$ rules out non-random strings (such as `1010...10') but permits their random-looking permutations. Still, there is a ``nearby'' credence function that is exchangeable. \x{Indeed, let $\widetilde{Cr} = Cr(\cdot \mid Freq_{\epsilon})$ result from conditioning on the frequency constraint alone. Since $Freq_{\epsilon}$ treats all reorderings alike, $\widetilde{Cr}$ remains exchangeable, and $Cr_{L^\star_{\epsilon,z}} = \widetilde{Cr}(\cdot \mid Rand_z)$. Being exchangeable, $\widetilde{Cr}$ is uniform within each frequency class $H_h$ (the strings with exactly $h$ heads). By a counting argument (Li and Vit\'anyi 2019,
Ex.~5.5.1(ii), p.~410), fewer than $2^{\log\binom{W}{h}-z} = \binom{W}{h}\,2^{-z}$ strings in $H_h$ violate $Rand_z$, less than a $2^{-z}$ fraction of every class. Averaging over the $H_h$ classes, $\widetilde{Cr}(\neg Rand_z) \leq 2^{-z}$. Conditioning on $Rand_z$ shifts credences by at most that amount: for
every $k \leq W$,
\begin{equation}
    |\widetilde{Cr}(x_1, \dots, x_k) - Cr_{L^\star_{\epsilon, z}}(x_1,
    \dots, x_k)| \leq 2^{-z}.
\end{equation}
With $z=\sqrt{W}$, at $W = 2^{100}$ the error is $2^{-2^{50}}$, which is vanishingly small.}

\textbf{Step 2.} We then use the Diaconis-Freedman theorem (1980b, Theorem 3) for finite exchangeable sequences to show that the exchangeable credence \x{$\widetilde{Cr}$} is  close to a mixture of iid models $\beta_{p}$ around $\beta_{0.5}$.  The difference is the $4k/W$ error from their result for binary sequences, where $k$ is the number of coordinates we are evaluating:
\begin{equation}
    |\x{\widetilde{Cr}}(x_1, \dots, x_k) - \int \beta_{p}(x_1, \dots, x_k) d\gamma(p)| \leq 4k/W.
\end{equation}
Since \x{$\widetilde{Cr}$ results from conditioning on $Freq_{\epsilon}$, it is concentrated on the frequency range}. By the Diaconis-Freedman theorem, the mixing measure  $\gamma(p)$ is induced by \x{$\widetilde{Cr}$}.  Hence,  $\gamma(p)$  assigns probability 1 to the range $p\in (0.5-\epsilon, 0.5+\epsilon)$. If $W$ is large (say, $2^{100}$) and we only look at a relatively small portion of it (say, $k=2^{10}$), $4k/W$ is close to 0 and the error is again tiny. 

\textbf{Step 3.} We compare $\beta_{0.5}$ with the mixture of iid  Bernoulli($p$) models with $p\in (0.5-\epsilon, 0.5+\epsilon)$. For any $p, q\in [0,1]$, we have $|\beta_p (x_1, \dots, x_k) - \beta_q (x_1, \dots, x_k) |\leq k|p-q|$. Taking averages does not make the bound worse than $k\epsilon$: 
\begin{equation}
    |\beta_{0.5}(x_1, \dots, x_k) - \int \beta_{p}(x_1, \dots, x_k) d\gamma(p)| \leq k\epsilon.
\end{equation}
If $W$ is large and we only look at a relatively small portion of it, the error is close to 0. For example, if $W=2^{100}$, $k=2^{10}$, then $k\epsilon\approx 2^{-37}$.

\textbf{Step 4.} The triangle inequality gives us the total bound as the sum of the three errors: 
\begin{equation}\label{totalbound}
    |Cr_{L^\star_{\epsilon, z}}(x_1, \dots, x_k) - \beta_{0.5}(x_1, \dots, x_k)| \leq \x{2^{-z}} + 4k/W + k\epsilon.
\end{equation}

This gives us an approximate version of the PP for the finite case. This proof extends to biased randomness too, as the fair-coin $\beta_{0.5}$ measure plays no special role. One can simply replace $\beta_{0.5}$ with $\beta_{\theta}$ throughout. The overall error bound retains the same form.

Notice that as $W$ grows, all three error terms in (\ref{totalbound}) shrink. In the transition to infinite histories ($W\rightarrow \infty$), we replace the finite constraint law  $L^\star_{\epsilon, z}$ with the infinite constraint law $L^\star$, yielding the exact PP proven in \S4. More precisely, if $z=z(W)\rightarrow \infty$ (e.g. $z=\sqrt{W}$), $\epsilon=\epsilon(W) \rightarrow 0$ (e.g. $\epsilon = \sqrt{\log W/W}$), and $k$ is fixed, then $\x{2^{-z}}\rightarrow 0$, $4k/W \rightarrow 0$, and $k\epsilon \rightarrow 0$. Hence, the finite bound (\ref{totalbound}) vanishes and we recover the exact PP. Allowing mixtures over exchangeability structures (as in \S5.1) only adds a model-uncertainty term, which likewise disappears in the same limit as the mixture concentrates on patterns compatible with $L^\star$. 

Recovering the exact infinite-histories result is evidence that the framework is correctly set up. It shows the infinite version is not a mere idealization but the precise and natural limit of the finite case, and thus all the more compelling.

\section{Implications}

The present framework has additional implications.

First, it shows that nomological dependence and probabilistic independence are compatible. The constraining law $L^\star$ imposes holistic, non-local dependencies on the entire infinite sequence. Yet, an agent with \x{exchangeable} credences who accepts this law must have credences that are perfectly independent (iid), just as the PP requires.\footnote{Roberts (2009) suggests that they are already compatible to an approximate degree, in the case of nomic frequentism. In our framework, the compatibility is exact.}

Second, \x{the framework connects global laws to single-case credences and to their empirical confirmation}. A central challenge for theories of chance based on long-run patterns is explaining how those patterns should inform an agent's credence about a single event, \x{and how such global patterns can be empirically assessed on the basis of finite observations}. Our framework provides a direct answer \x{to both}. An agent who accepts $L^\star$ and holds exchangeable credences will assign precisely the same single-case credences she would if she were committed to the standard probabilistic law $L$\x{, and her conditional credences supply the likelihoods by which finite evidence bears on rival $\star$-laws (\S2)}. \x{Under $L$, this coordination between credence and objective probability is standardly secured by positing the PP; under $L^\star$, it can be derived (\S3.3).}

Finally, the derivation of the PP is metaphysically neutral, undermining the suggestion that only a Humean account could justify the PP. Our argument for the PP does not depend on whether one adopts Humeanism or non-Humeanism. 
What the PP requires is a balance between metaphysics and epistemology. $L^\star$ laws find a ``sweet spot'' by embedding enough structure into the metaphysics to make the epistemology straightforward.

\section{Conclusion}

We have shown that the Principal Principle naturally emerges when algorithmic randomness meets exchangeable credences. The present framework unites concepts developed independently---de Finetti's exchangeability, Martin-L\"of randomness, chance-credence norms, and statistical constraining laws. The framework extends to partial exchangeability and finite histories, where error bounds vanish in the infinite limit. In this framework, the PP need not be a primitive posit. Instead, it follows mathematically from the alignment between how laws constrain patterns and how rational agents learn them---an alignment that holds regardless of whether one adopts a Humean or non-Humean metaphysics of laws. The result delivers what realists about chance have long sought: a non-circular justification of the chance-credence norm. More broadly, the framework opens new paths for understanding how probability simultaneously describes physical reality and constrains rational belief.

\section*{Acknowledgement} 
We thank the editors and two anonymous reviewers at \emph{The British Journal for the Philosophy of Science} for valuable feedback. We are also grateful for helpful discussions with Marshall Abrams, Emily Adlam, Boris Babic, Jacob Barandes, Craig Callender, Susumu Cato, Harry Crane, Dmitri Gallow, Veronica Gomez Sanchez, Alan H\'ajek, Christopher Hitchcock, Mario Hubert, Carl Hoefer, Thomas Icard,  Mahmoud Jalloh, Gabrielle Kerbel, Alexander Kocurek, Barry Loewer, Kelvin McQueen, Krzysztof Mierzewski,  Wayne Myrvold, Joseph Obrien, Jun Otsuka, Tomasz Placek, Ezra Rubenstein, Wolfgang Schwarz, Charles Sebens, Glenn Shafer, Shelly Yiran Shi, Brian Skyrms, Tatsuya Yoshii,  David Wallace,  Isaac Wilhelm,  Francesca Zaffora Blando, Jiji Zhang, Snow Zhang, and audiences at 2024 Philosophy of Science Association Biennial Meeting,  University of Tokyo,  University of Hong Kong,  University of California Berkeley, Chapman University, CalTech, Jagiellonian University,  University of California San Diego, and Stanford University. 
The authors used several versions of ChatGPT, Codex, and Claude in discussion, to check the proofs, and for editorial assistance. Codex suggested an elegant simplification of step 1 of our finite history proof. Any remaining errors are ours.
EKC is supported by Grant 63209 from the John Templeton Foundation. The opinions expressed in this publication are those of the authors and do not necessarily reflect the views of those kind enough to talk with us or the John Templeton Foundation.
 


\begin{center}
\large{Bibliography}
\end{center}

\vspace{.25cm}
\noindent
Barrett, Jeffrey A. and Chen, Eddy Keming. (2025) “Algorithmic Randomness and Probabilistic Laws,” \emph{The British Journal for the Philosophy of Science}, forthcoming. \\ \href{https://doi.org/10.1086/738573}{https://doi.org/10.1086/738573} \ \href{https://arxiv.org/abs/2303.01411}{https://arxiv.org/abs/2303.01411}

\vspace{.25cm}
\noindent
Belot, Gordon. (2024) “Unprincipled,” \emph{The Review of Symbolic Logic}, \textbf{17}(2), 435-474. \\
\href{https://doi.org/10.1017/S1755020323000151}{https://doi.org/10.1017/S1755020323000151}

\vspace{.25cm}
\noindent
Belot, Gordon. (2023) “That Does Not Compute: David Lewis on Credence and Chance,” \emph{Philosophy of Science}, \textbf{90}(5), pp. 1130-1139. \ \href{https://doi.org/10.1017/psa.2023.11}{https://doi.org/10.1017/psa.2023.11}

\vspace{.25cm}
\noindent
Cariani, Fabrizio. (2017) ``Chance, Credence and Circles,'' \emph{Episteme}, \textbf{14(1)}, 49–58. \\ \href{https://doi.org/10.1017/epi.2016.48}{https://doi.org/10.1017/epi.2016.48}

\vspace{.25cm}
\noindent
Chen, Eddy Keming (2022) ``Fundamental Nomic Vagueness,''  \emph{The Philosophical Review}, 131 (1): 1–49. \href{https://doi.org/10.1215/00318108-9415127}{https://doi.org/10.1215/00318108-9415127}

\vspace{.25cm}
\noindent
Chen, Eddy Keming and Sheldon Goldstein (2022) ``Governing without a Fundamental Direction of Time: Minimal Primitivism about Laws of Nature,'' in Yemima Ben-Menahem (ed.), \emph{Rethinking the Concept of Law of Nature}, Springer, pp.21-64. 

\noindent
\href{https://arxiv.org/abs/2109.09226}{https://arxiv.org/abs/2109.09226}

\vspace{.25cm}
\noindent
Chen, Eddy Keming (2024a) ``The Wentaculus: Density Matrix Realism Meets the Arrow of Time,'' in  A. Bassi, S. Goldstein, R. Tumulka, and N. Zangh\`i (eds.), \emph{Physics and the Nature of Reality: Essays in Memory of Detlef D\"urr}, Heidelberg: Springer, pp. 87-104.

\vspace{.25cm}
\noindent
\x{Chen, Eddy Keming (2024b) \emph{Laws of Physics}, Cambridge University Press.}

\vspace{.25cm}
\noindent
Dasgupta, Abhijit (2011) ``Mathematical Foundations of Randomness,'' in Prasanta Bandyopadhyay and Malcolm Forster (eds.), \emph{Philosophy of Statistics (Handbook of the Philosophy of Science: Volume 7)}, Amsterdam: Elsevier, pp. 641–710. 

\vspace{.25cm}
\noindent
de Finetti, Bruno. (1937) “La pr\'evision: ses lois logiques, ses sources subjectives,” \emph{Annales de l’Institut Henri Poincar\'e}, \textbf{7}(1), 1–68.

\vspace{.25cm}
\noindent
de Finetti, Bruno. (1974–1975) \emph{Theory of Probability}, Vols. 1 and 2. Trans. A. Machi and A. F. M. Smith. New York: Wiley.

\vspace{.25cm}
\noindent
Diaconis, Persi and Freedman, David. (1980a) ``De Finetti's Theorem for Markov Chains,'' \emph{Annals of Probability}, \textbf{8}(1), pp.115-130. 

\vspace{.25cm}
\noindent
Diaconis, Persi and Freedman, David. (1980b) ``Finite Exchangeable Sequences,'' \emph{Annals of Probability}, \textbf{8}(4), pp.745–764.

\vspace{.25cm}
\noindent
Diaconis, Persi and Brian Skyrms (2018) \textit{Ten Great Ideas about Chance}, Princeton University Press: Princeton and Oxford.

\vspace{.25cm}
\noindent
Eagle, Antony, (2021) ``Chance versus Randomness,'' \emph{The Stanford Encyclopedia of Philosophy}, (Spring 2021 Edition), Edward N. Zalta (ed.), \\ \href{https://plato.stanford.edu/archives/spr2021/entries/chance-randomness/}{https://plato.stanford.edu/archives/spr2021/entries/chance-randomness/}

\vspace{.25cm}
\noindent
Elga, Adam (2004) ``Infinitesimal Chances and the Laws of Nature,'' \emph{Australasian Journal of Philosophy}, 82(1), pp.67-76. 

\vspace{.25cm}
\noindent
Freedman, David A. (1962) ``Invariants under Mixing which Generalize de Finetti's Theorem,'' \emph{Annals of Mathematical Statistics}, \textbf{33}(3), pp.916–923.

\vspace{.25cm}
\noindent
Goldstein, Sheldon (2025) ``Bohmian Mechanics,'' \emph{The Stanford Encyclopedia of Philosophy}, (Fall 2025 Edition), Edward N. Zalta \& Uri Nodelman (eds.), \\ \href{https://plato.stanford.edu/archives/fall2025/entries/qm-bohm/}{https://plato.stanford.edu/archives/fall2025/entries/qm-bohm/}

\vspace{.25cm}
\noindent
Hall, Ned (1994) ``Correcting the Guide to Objective Chance,'' \emph{Mind} 103: 504-517. 

\vspace{.25cm}
\noindent
H\'ajek, Alan. (2009) ``Fifteen Arguments Against Hypothetical Frequentism,'' \emph{Erkenntnis}, \textbf{70}, pp.211–235.

\vspace{.25cm}
\noindent
H\'ajek, Alan (2023) "Interpretations of Probability", in Edward N. Zalta \& Uri Nodelman (eds.) \emph{The Stanford Encyclopedia of Philosophy (Winter 2023 Edition)}. 

\noindent
\href{https://plato.stanford.edu/archives/win2023/entries/probability-interpret/}{https://plato.stanford.edu/archives/win2023/entries/probability-interpret/}

\vspace{.25cm}
\noindent
Henderson, Leah. (2024) “The Problem of Induction,” \emph{The Stanford Encyclopedia of Philosophy}. \ \href{https://plato.stanford.edu/archives/sum2024/entries/induction-problem/}{https://plato.stanford.edu/archives/sum2024/entries/induction-problem/}


\vspace{.25cm}
\noindent
Hume, David. (1748) \emph{An Enquiry Concerning Human Understanding}. London: A. Millar.


\vspace{.25cm}
\noindent
Kotzen, Matthew. (2018) ``Comments on Richard Pettigrew's \emph{Accuracy and the Laws of Credence},'' \emph{Philosophy and Phenomenological Research}, \textbf{96(3)}, 776–783. \\ \href{https://doi.org/10.1111/phpr.12504}{https://doi.org/10.1111/phpr.12504}

\vspace{.25cm}
\noindent
Lewis, David. (1980) ``A Subjectivist’s Guide to Objective Chance,'' in Richard C. Jeffrey (ed.), \emph{Studies in Inductive Logic and Probability}, Vol. II, Berkeley: University of California Press, 263–293.

\vspace{.25cm}
\noindent
Lewis, David (1994) ``Humean Supervenience Debugged'', \emph{Mind} 103:473-490

\vspace{.25cm}
\noindent
Li, Ming and Paul Vit\'anyi (2019). \emph{An Introduction to Kolmogorov Complexity and Its Applications, 4th Edition}, New York, NY: Springer. 

\vspace{.25cm}
\noindent
Lin, Hanti. (2024) “Bayesian Epistemology,” \emph{The Stanford Encyclopedia of Philosophy} (2024). \ \href{https://plato.stanford.edu/archives/sum2024/entries/epistemology-bayesian/}{https://plato.stanford.edu/archives/sum2024/entries/epistemology-bayesian/}

\vspace{.25cm}
\noindent
Loewer, Barry (2004) ``David Lewis’s Humean Theory of Objective Chance'', \emph{Philosophy of Science} 71(5):1115-25

\vspace{.25cm}
\noindent
Loewer, Barry (2020) ``The Mentaculus Vision'', In Valia Allori (ed.), \emph{Statistical Mechanics and Scientific Explanation: Determinism, Indeterminism and Laws of Nature}, Singapore: World Scientific, pp. 3-29.

\vspace{.25cm}
\noindent
Martin-L\"of, Per (1966) ``The definition of random sequences,'' \emph{Information and Control} 9(6):602--619.


\vspace{.25cm}
\noindent
Mierzewski, Krzysztof and Zaffora Blando, Francesca  (2024) ``Two Great Ideas about Chance,'' symposium presentation at the 2024 Philosophy of Science Association Biennial Meeting

\vspace{.25cm}
\noindent
Miller, David. (1966) ``A Paradox of Information,'' \emph{The British Journal for the Philosophy of Science}, \textbf{17}(1), pp.59–61.

\vspace{.25cm}
\noindent
\x{Nies, Andr\'e. (2009) \emph{Computability and Randomness}. Oxford: Oxford University Press.}

\vspace{.25cm}
\noindent
Pettigrew, Richard. (2012) ``Accuracy, Chance, and the Principal Principle,'' \emph{The Philosophical Review}, \textbf{121(2)}, 241–275. 
\href{https://doi.org/10.1215/00318108-1539098}{https://doi.org/10.1215/00318108-1539098}

\vspace{.25cm}
\noindent
Pettigrew, Richard. (2016) \emph{Accuracy and the Laws of Credence}, Oxford: Oxford University Press.

\vspace{.25cm}
\noindent
\x{Porter, Christopher P. (2016) ``On Analogues of the Church--Turing Thesis in Algorithmic Randomness,'' \emph{The Review of Symbolic Logic}, \textbf{9}(3), pp.~456--479.} \\
\href{https://doi.org/10.1017/S1755020316000113}{https://doi.org/10.1017/S1755020316000113}

\vspace{.25cm}
\noindent
Porter, Christopher P. (2020) ``Biased Algorithmic Randomness'' in \emph{Algorithmic Randomness: Progress and Prospects},
Edited by Johanna N.\ Y.\ Franklin and Christopher P.\ Porter. Lecture Notes in Logic, 50, Association for Symbolic Logic, pp. 206-231.

\vspace{.25cm}
\noindent
\x{Porter, Christopher P. (2021) ``The Equivalence of Definitions of Algorithmic Randomness,'' \emph{Philosophia Mathematica}, \textbf{29}(2), pp.~153--194.} \\
\href{https://doi.org/10.1093/philmat/nkaa039}{https://doi.org/10.1093/philmat/nkaa039}

\vspace{.25cm}
\noindent
Roberts, John T. (2009) ``Laws about Frequencies'' manuscript,  \\ \href{https://philpapers.org/rec/ROBLAF}{https://philpapers.org/rec/ROBLAF}.


\vspace{.25cm}
\noindent
Schwarz, Wolfgang. (2014) ``Proving the Principal Principle,'' in Alastair Wilson (ed.), \emph{Chance and Temporal Asymmetry}, Oxford: Oxford University Press, pp. 81-99.

\vspace{.25cm}
\noindent
Shafer, Glenn. (2023) ```That’s what all the old guys said.'
The many faces of Cournot’s principle,'' Working Paper 60,  \href{https://www.probabilityandfinance.com/articles/60.pdf}{https://www.probabilityandfinance.com/articles/60.pdf}. 

\vspace{.25cm}
\noindent
Skyrms, Brian. (1977) “Resiliency, Propensities, and Causal Necessity,” \emph{The Journal of Philosophy}, \textbf{74}(11), pp.704–713.

\vspace{.25cm}
\noindent
Skyrms, Brian. (1980) \emph{Causal Necessity: A Pragmatic Investigation of the Necessity of Laws}. New Haven: Yale University Press.

\vspace{.25cm}
\noindent
Skyrms, Brian. (2014) “Grades of Inductive Skepticism,” \emph{Philosophy of Science}, \textbf{81}(3), pp.303–312. \ \href{https://doi.org/10.1086/676637}{https://doi.org/10.1086/676637}

\vspace{.25cm}
\noindent
Strevens, Michael. (2017) “Notes on Bayesian Confirmation Theory,” manuscript,  \ \href{https://www.strevens.org/bct/BCT.pdf}{https://www.strevens.org/bct/BCT.pdf}

\vspace{.25cm}
\noindent
Thau, Michael (1994) ``Undermining and Admissibility'', \emph{Mind} 103: 491-503

\vspace{.25cm}
\noindent
Zabell, Sandy L. (2005) \emph{Symmetry and Its Discontents: Essays on the History of Inductive Probability}. Cambridge: Cambridge University Press.

\vspace{.15cm}
\noindent
Zaffora Blando, Francesca. (2025) ``Bayesian Merging of Opinions and Algorithmic Randomness,'' \emph{The British Journal for the Philosophy of Science}, \textbf{76}(4), pp. 921-952. \\ \href{https://doi.org/10.1086/721758}{https://doi.org/10.1086/721758}

\end{document}